\documentclass[twocolumn,floatfix,aps,prb]{revtex4}
\usepackage{graphicx}
\usepackage{amsmath}
\usepackage{amssymb}
\usepackage{bm}

\begin{document}

\title{Even-odd flux quanta effect in the Fraunhofer oscillations of an edge-channel Josephson junction}
\author{B. Baxevanis}
\affiliation{Instituut-Lorentz, Universiteit Leiden, P.O. Box 9506, 2300 RA Leiden, The Netherlands}
\author{V. P. Ostroukh}
\affiliation{Instituut-Lorentz, Universiteit Leiden, P.O. Box 9506, 2300 RA Leiden, The Netherlands}
\author{C. W. J. Beenakker}
\affiliation{Instituut-Lorentz, Universiteit Leiden, P.O. Box 9506, 2300 RA Leiden, The Netherlands}
\date{November 2014}
\begin{abstract}
We calculate the beating of $h/2e$ and $h/e$ periodic oscillations of the flux-dependent critical supercurrent $I_c(\Phi)$ through a quantum spin-Hall insulator between two superconducting electrodes. A conducting pathway along the superconductor connects the helical edge channels via a non-helical channel, allowing an electron incident on the superconductor along one edge to be Andreev reflected along the opposite edge. In the limit of small Andreev reflection probability the resulting even-odd effect is described by $I_c\propto|\cos(e\Phi/\hbar)+f|$, with $|f|\ll 1$ proportional to the probability for phase-coherent inter-edge transmission. Because the sign of $f$ depends on microscopic details, a sample-dependent inversion of the alternation of large and small peaks is a distinctive feature of the beating mechanism for the even-odd effect.
\end{abstract}
\maketitle

\begin{figure}[tb]
\centerline{\includegraphics[width=0.6\linewidth]{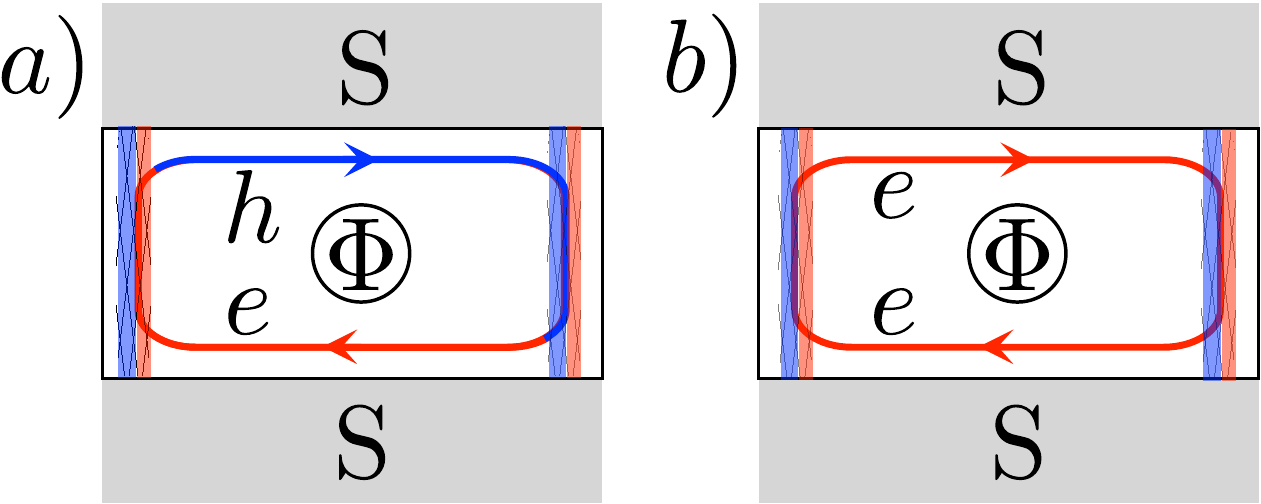}}
\caption{Beating mechanism for the even-odd effect in the Fraunhofer oscillations. For uncoupled edges the flux periodicity is $h/2e$, corresponding to the transfer of a charge-$2e$ Cooper pair along the left or right edge channel (blue/red hatched strips). The edge channels are coupled by a conducting path along the NS interface, allowing for a circulating loop of charge $\pm e$ with $h/e$ flux periodicity. The circulating loop may be partly $e$-type (red lines) and partly $h$-type (blue), as in panel \textit{a}, or it may be entirely of one charge-type (entirely $e$, as in panel \textit{b}, or entirely $h$). Both loops contribute to the even-odd effect, but panel \textit{a} dominates when the Andreev reflection probability $\Gamma$ is small.  (It is of order $\Gamma$, while panel \textit{b} is of order $\Gamma^2$.)
}
\label{fig_intro}
\end{figure}

Superconductor--normal-metal--superconductor (SNS) junctions with edge channel conduction in the normal region are governed by the interplay of charge $e$ and charge $2e$ transport: Charge can only enter or exit the superconductor in units of $2e$, but in the normal region this Cooper pair can be split over opposite edges, when an electron incident on the NS interface along one edge is Andreev reflected as a hole along the opposite edge.

For quantum Hall edge channels this mechanism produces Fraunhofer oscillations (oscillations of the critical current with enclosed flux $\Phi$) having a fundamental period of $h/e$, twice the usual periodicity.\cite{Ost11} These are chiral edge channels, so Andreev reflection along the edge of incidence is forbidden and only the circulating path of Fig.\ \ref{fig_intro}a contributes to the supercurrent. When the edge channels allow for propagation in both directions, the critical current includes the usual $h/2e$-periodic contributions from Andreev reflection along a single edge, and further $h/e$ periodic contributions from circulating paths without charge transfer (Fig.\ \ref{fig_intro}b).

Here we investigate this beating of $h/e$ and $h/2e$ periodic contributions to the Fraunhofer oscillations. We are motivated by recent work on proximity induced superconductivity in quantum spin-Hall (QSH) insulators,\cite{Kne12,Har13,Pri14,Shi14,Lee14,Hui14,Tka14} which in one series of experiments\cite{Pri14} showed Fraunhofer oscillations with an even-odd effect: Large peaks in the critical current at even multiples of $h/2e$ alternate with smaller peaks at odd multiples.

The QSH insulator has helical edge channels (with direction of motion tied to the spin), so we consider that case in what follows (although the beating mechanism for the even-odd effect does not rely on helicity). Following Ref.\ \onlinecite{Lee14} we assume that the superconductors dope the contacted QSH insulator, locally pushing the Fermi level in the conduction band. The broad conducting pathway that appears along the NS interface will be gapped by the superconducting proximity effect, but a narrow gapless channel may remain because superconductivity only becomes effective at some penetration length $\xi_0$ from the NS interface. (Ref.\ \onlinecite{Pri14} estimates $\xi_0\gtrsim 240\,{\rm nm}$, comparable to the estimated width of the edge states.) This channel provides a connection between the helical edge states that is non-helical, meaning that either spin can propagate in both directions.

To describe the phase-coherent coupling of helical and non-helical edge channels we study a network model of the Josephson junction, inspired by the spectral theory of graphs\cite{Smi09} and as a counterpart to network models of the quantum Hall effect.\cite{Cha88,Kra05} As we will show, all information on the temperature and flux dependence of the supercurrent can be encoded in the product of a permutation matrix, representing the connectivity of the network, and a block-diagonal matrix describing the relation between incoming and outgoing modes at each node of the network.

\begin{figure}[tb]
\centerline{\includegraphics[width=1\linewidth]{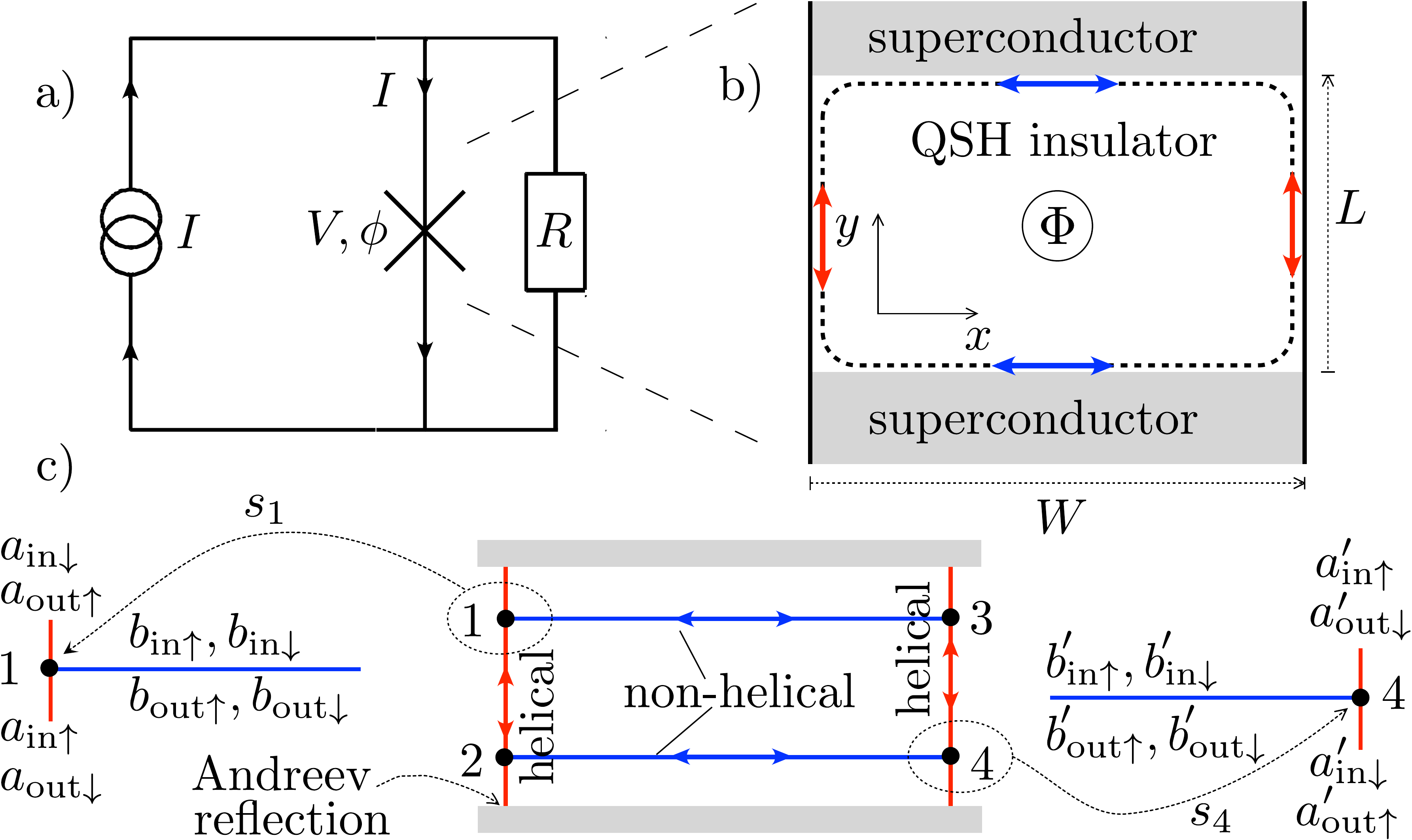}}
\caption{Josephson junction in a current-biased circuit (panel a), to study the dependence of the critical current $I_c$ on the magnetic flux $\Phi$ enclosed by a circulating edge channel (panel b). The network model of the Josephson junction is illustrated in panel c. Helical modes (red, amplitudes $a_\uparrow,a_\downarrow$) and non-helical modes (blue, amplitudes $b_\uparrow,b_\downarrow$) are coupled at four nodes by a scattering matrix $s_n$, relating incoming and outgoing amplitudes.
}
\label{fig_layout}
\end{figure}

{\em Edge-channel Josephson junction ---}
We consider the Josephson junction geometry of Fig.\ \ref{fig_layout}a. A current $I$ is passed between two superconducting electrodes at phase difference $\phi$, related to the voltage $V$ over the junction by the Josephson relation $d\phi/dt=(2e/\hbar)V$. Upon increasing the current bias, the junction switches from zero to finite {\sc dc} voltage at a critical current $I_c$, dependent on the enclosed magnetic flux $\Phi$. If phase fluctuations can be neglected (for a low-impedance environment), the critical current is given by
\begin{equation}
I_c(\Phi)=\textstyle{\max_{\phi}}\,|I(\phi,\Phi)|.\label{Icdef}
\end{equation}
We seek the oscillatory $\Phi$-dependence of $I_c$ (Fraunhofer oscillations) in a junction where the current flows along the edges, rather than through the bulk.

Referring to Fig.\ \ref{fig_layout}b, the junction has width $W$ (edges at $x=0,W$) and length $L$ (normal-superconductor or NS interfaces at $y=0,L$). We choose a gauge where the superconducting pair potential $\Delta_0$ is real. A vector potential $\bm{A}=A_y\hat{y}$ in the $y$-direction,
\begin{equation}
A_y=\frac{\Phi x}{LW}+\frac{\Phi_0\phi}{2\pi}\delta(y-L/2),\;\;\Phi_0\equiv\frac{h}{2e},\label{Adef}
\end{equation}
then accounts for the phase difference between the NS interfaces.

{\em Network model ---}
To capture the essence of the problem, while still allowing for analytical solution, we represent the scattering processes by a network (Fig.\ \ref{fig_layout}c). At the nodes $n=1,2,3,4$ the helical edge channels along $x=0,W$ are coupled to a single-mode non-helical channel along $y=0,L$. Each node has a $4\times 4$ electronic scattering matrix $s_n$, which relates incoming and outgoing wave amplitudes of the helical channel, $a=(a_\uparrow,a_\downarrow)$, and the non-helical channel, $b=(b_\uparrow,b_\downarrow)$, according to
\begin{equation}
\begin{pmatrix}
a\\
b
\end{pmatrix}_{\rm out}=s_n\begin{pmatrix}
a\\
b
\end{pmatrix}_{\rm in}.\label{sndef}
\end{equation}
The short-range scattering at a node can be taken as energy-independent, so the hole scattering matrix is simply the complex conjugate $s_n^\ast$. We collect these matrices in the unitary matrix $s_{\rm node}=s_1\oplus s_1^\ast\oplus \cdots\oplus s_4\oplus s_4^\ast$, consisting of eight $4\times 4$ blocks arranged along the diagonal.

Since the effect of the magnetic field is only felt on long length scales, we can assume that $s_n$ preserves time-reversal symmetry. The requirement
\begin{equation}
s_n=\begin{pmatrix}
\sigma_y&0\\
0&\sigma_y
\end{pmatrix}s_n^{\rm T}\begin{pmatrix}
\sigma_y&0\\
0&\sigma_y
\end{pmatrix},\label{TRS}
\end{equation}
together with unitarity, $s_n^\dagger s_n=1$, imposes the form\cite{Jia13}
\begin{equation}
s_n=\begin{pmatrix}
e^{2i\psi_n}\sigma_0\sqrt{\Gamma_n}&e^{i\psi_n+i\psi'_n}U_n\sqrt{1-\Gamma_n}\\
e^{i\psi_n+i\psi'_n}U^{\dagger}_n\sqrt{1-\Gamma_n}&-e^{2i\psi'_n}\sigma_0\sqrt{\Gamma_n}
\end{pmatrix}.\label{snUn}
\end{equation}
Helical and non-helical channels are coupled with probability $1-\Gamma_n$, while $U_n\in{\rm SU}(2)$ describes the spin-mixing associated with that coupling. [Eq.\ \eqref{TRS} is satisfied because $\sigma_y U_n^{\rm T}\sigma_y=U_n^\dagger$ for any ${\rm SU}(2)$ matrix $U(n)$.] Time-reversal symmetry forbids spin mixing within the helical or non-helical channel, which is why the upper-left and lower-right blocks of $s_n$ are proportional to the $2\times 2$ unit matrix $\sigma_0$.

The nodes are connected by a unitary bond matrix $s_{\rm bond}$, which is the product of a diagonal matrix of phase factors and a permutation matrix. We decompose $s_{\rm bond}=s_{\rm left}\oplus s_{\rm right}\oplus s_{\rm bottom}\oplus s_{\rm top}$ in terms of matrices $s_{\rm left}$ and $s_{\rm right}$ that connect the $a$-amplitudes (along $x=0$ and $x=W$, with phase factor $e^{i\varepsilon L/\hbar v}\exp[i\tau_z (e/\hbar)\int A_y\,dy]$) and matrices $s_{\rm bottom}$ and $s_{\rm top}$ that connect the $b$-amplitudes (along $y=0$ and $y=L$, with phase factor $e^{i\varepsilon W/\hbar v}$).  Andreev reflection is included in $s_{\rm left}$ and $s_{\rm right}$ via matrix elements that connect a node to itself, switching electron-hole and spin-band with phase factor
\begin{equation}
s_{\rm A}=i\alpha\tau_{y}\otimes\sigma_y,\;\;\alpha(\varepsilon)=i\varepsilon/\Delta_0+\sqrt{1-\varepsilon^2/\Delta_0^2}.\label{sAdef}
\end{equation}
(The Pauli matrices $\tau_i$ and $\sigma_i$ act, respectively, on the electron-hole $e,h$ and spin $\uparrow,\downarrow$ degree of freedom.)

Knowledge of $s_{\rm node}$ and $s_{\rm bond}$ determines the entire spectrum of the network.\cite{Smi09} A bound state at energy $|\varepsilon|<\Delta_0$ corresponds to a unit eigenvalue of $M(\varepsilon)=s_{\rm node}s_{\rm bond}(\varepsilon)$, leading to the determinantal equation ${\rm Det}\,[1-M(\varepsilon)]=0$. The density of states of the continuous spectrum at $|\varepsilon|>\Delta_0$ is given by\cite{appendix}
\begin{equation}
\rho(\varepsilon)=-\frac{1}{\pi}\frac{d}{d\varepsilon}\,{\rm Im}\,\ln\,{\rm Det}[1-M(\varepsilon+i0^+)]+{\rm constant},\label{rhodef}
\end{equation}
where the ``constant'' refers to $\phi$-independent terms. The Josephson current at temperature $T$ then follows from\cite{Bee91,Bro97}
\begin{align}
I(\phi,\Phi)&=-kT\frac{2e}{\hbar}\frac{d }{d \phi}\sum_{p=0}^\infty\ln {\rm Det}\, [1-M(i\omega_p)], \label{IMatsubara}
\end{align}
as a sum over fermionic Matsubara frequencies $\omega_p=(2p+1)\pi kT$. This expression assumes that the system equilibrates without restrictions on the fermion parity, so it holds on time scales long compared to the quasiparticle poisoning time (otherwise there would appear an additional sum over bosonic Matsubara frequencies).\cite{Bee13}

{\em Uncoupled edges ---} When $kT\gg\hbar v/W$ there is no phase-coherent coupling between the edges at $x=0$ and $x=W$. We may then set $s_{\rm top}$ and $s_{\rm bottom}$ to zero in the evaluation of the determinant in Eq.\ \eqref{IMatsubara}, with the result
\begin{align}
&I(\phi,\Phi)=I_{\rm edge}(\phi)+I_{\rm edge}(\phi+2\pi\Phi/\Phi_0),\label{IIleftIright}\\
&I_{\rm edge}(\phi)=kT\frac{4e}{\hbar}\sin\phi\sum_{p=0}^\infty[2\cos\phi+\zeta(\omega_p)+1/\zeta(\omega_p)]^{-1},\nonumber\\
&\zeta(\omega)=\Gamma^2 e^{-2\omega L/\hbar v}\left[\sqrt{1+\omega^2/\Delta_0^2}-\omega/\Delta_0\right]^2.\label{zetadef}
\end{align}
(To simplify the formulas we have taken identical $\Gamma_n\equiv\Gamma$.) 

For $\Gamma\rightarrow 1$ we recover the short-junction-to-long-junction crossover formula of Ref.\ \onlinecite{Bee13}, which in the short-junction limit $L\ll\hbar v/\Delta_0$ and for low temperatures $kT\ll\Delta_0$ results in a critical current
\begin{equation}
I_c(\Phi)=\frac{e\Delta_0}{2\hbar}\bigl(1+|\cos(\pi\Phi/\Phi_0)|\bigr)\label{Icshortlow}
\end{equation}
with minima that are offset from zero, in agreement with Ref.\ \onlinecite{Lee14}. For $\Gamma\ll 1$, still in the short-junction and low-temperature limit, we find instead
\begin{align}
&I(\phi,\Phi)=I_0\sin(\phi+\pi\Phi/\Phi_0)\cos(\pi\Phi/\Phi_0)\label{IshortlowsmallGamma}\\
&\Rightarrow I_c(\Phi)=I_0|\cos(\pi\Phi/\Phi_0)|,\;\;I_0=\frac{8e\Delta_0}{3\pi\hbar}\Gamma^2. \label{Icshortlow2}
\end{align}
For these uncoupled edges the critical current is $h/2e$ periodic in $\Phi$.

{\em Coupled edges ---} The effect on the supercurrent of a phase-coherent coupling of the edges can be studied perturbatively in powers of $e^{-\pi kT W/\hbar v}$, by expanding the logarithmic determinant in Eq.\ \eqref{IMatsubara} with the help of the formula
\begin{align}
&\ln{\rm Det}\,(1-M_0-\delta M)=\ln{\rm Det}\,(1-M_0)\nonumber\\
&\quad-\sum_{n=1}^{\infty}\frac{1}{n}{\rm Tr}\,\bigl[(1-M_0)^{-1}\delta M\bigr]^{n}.\label{Detexpansion}
\end{align}
The lowest order contribution with $h/e$ periodicity in $\Phi$ is given by
\begin{align}
&\delta I_{h/e}=kT\frac{2e}{\hbar}\frac{d }{d \phi}\,{\rm Tr}\,s_{\rm node}(1-s_{\rm left}s_{\rm node})^{-1}s_{\rm top}s_{\rm node}\nonumber\\
&\cdot (1-s_{\rm right}s_{\rm node})^{-1}s_{\rm bottom}\bigl|_{\varepsilon=i\omega_0} + \, \{s_{\rm left}\leftrightarrow s_{\rm right}\},
\end{align}
describing a quasiparticle that encircles the junction clockwise or anti-clockwise. 

The effect of this contribution is largest for small Andreev reflection probability $\Gamma_n\ll 1$. To first order in $\Gamma$, and in the low-temperature, short-junction limit, we find
\begin{align}
\delta I_{h/e}={}&(8e/\hbar)kTe^{-2\pi kTW/\hbar v}\sin(\phi+\pi\Phi/\Phi_0)\nonumber\\
&\times \bigl(\sqrt{\Gamma_1\Gamma_2}+\sqrt{\Gamma_1\Gamma_4}+\sqrt{\Gamma_3\Gamma_4}+\sqrt{\Gamma_3\Gamma_2}\bigr)\nonumber\\
&\times \sin(\gamma_{2}-\gamma_4)\sin(\gamma_1-\gamma_3).\label{Ihoveregamma}
\end{align}
[To simplify a lengthy general expression we made a definite choice $U_n=e^{i\gamma_n\sigma_x}$, $\psi_n=\psi'_n=0$ for the spin-mixing matrices.] Without spin mixing, for $\gamma_n=0$, the contribution \eqref{Ihoveregamma} of order $\Gamma$ vanishes, but there is a nonzero contribution of order $\Gamma^2$,
\begin{align}
\delta I_{h/e}={}&(8e/\hbar)kTe^{-2\pi kT W/\hbar v}\bigl[(\sin(\phi-\pi\Phi/\Phi_0)\Gamma_1\Gamma_2\nonumber\\
&+\sin(\phi+3\pi\Phi/\Phi_0)\Gamma_3\Gamma_4)\bigr].\label{Ihoveregamma2}
\end{align}
The contributions \eqref{Ihoveregamma} and \eqref{Ihoveregamma2} correspond to the pathways show in Figs.\ \ref{fig_intro}a and \ref{fig_intro}b, respectively.

\begin{figure}[tb]
\centerline{\includegraphics[width=0.8\linewidth]{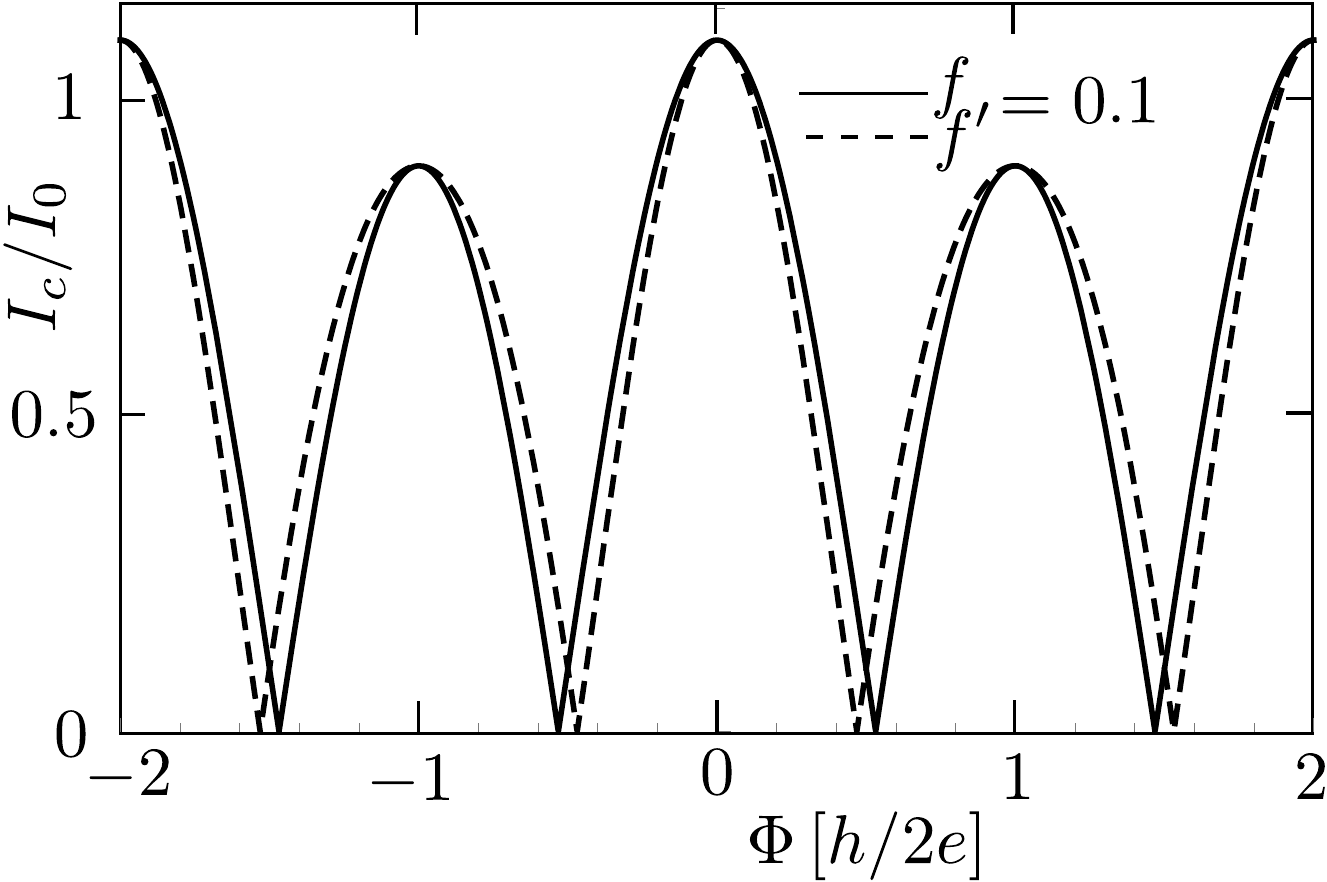}}
\caption{Even-odd effect in the Fraunhofer oscillations of the critical current due to the beating of $h/e$ and $h/2e$ oscillations. The curves are calculated with spin mixing from Eq.\ \eqref{Iresult1} (solid lines, dominated by the path of Fig.\ \ref{fig_intro}a) and without spin mixing from Eq.\ \eqref{Iresult2} (dashed lines, dominated by the path of Fig.\ \ref{fig_intro}b).
}
\label{fig_evenodd}
\end{figure}

Addition of $\delta I_{h/e}$ to the zeroth order supercurrent \eqref{IshortlowsmallGamma} (for identical $\Gamma_n\equiv\Gamma$) gives the critical current
\begin{subequations}
\label{Iresult1}
\begin{align}
&I_c(\Phi)=I_0|\cos(\pi\Phi/\Phi_0)+f|,\label{Ireuslt1a}\\
&f=\frac{12\pi kT}{\Delta_0\Gamma}e^{-2\pi kTW/\hbar v}\sin(\gamma_{2}-\gamma_4)\sin(\gamma_1-\gamma_3),\label{Iresult1b}
\end{align}
\end{subequations}
with spin mixing at the nodes, and
\begin{subequations}
\label{Iresult2}
\begin{align}
&I_c(\Phi)=I_0|\cos(\pi\Phi/\Phi_0)+f'\cos(2\pi\Phi/\Phi_0)|,\label{Iresult2a}\\
&f'=(6\pi kT/\Delta_0)e^{-2\pi kTW/\hbar v},\label{Iresult2b}
\end{align}
\end{subequations}
without spin mixing. Both types of Fraunhofer oscillations are $h/e$ periodic, with an even-odd effect of relative magnitude $f$ or $f'$, see Fig.\ \ref{fig_evenodd}.

{\em Comparison with experiment ---} Turning now to the experiment that motivated this analysis,\cite{Pri14} we first of all notice that the observed even-odd effect appears already for the first few peaks around zero field. An explanation in terms of a Lorentz-force induced asymmetry in the current distribution is therefore unlikely.\cite{note1,Led99,Bar99,Hei98,Har02} The $h/e$-periodic Josephson effect of Majorana zero-modes\cite{Fu09} is spoiled, on the time scale of the experiment, by any small amount of quasiparticle poisoning,\cite{Lee14} so an explanation along these lines is not viable. A conducting pathway through the bulk, parallel to the edges, can explain the data\cite{Pri14} --- but only if it is located within 10\% of the device center (the flux $\Phi$ needs to be accurately partitioned into twice $\Phi/2$). The mechanism proposed here does not require any such fine tuning.

\begin{figure}[tb]
\centerline{\includegraphics[width=0.9\linewidth]{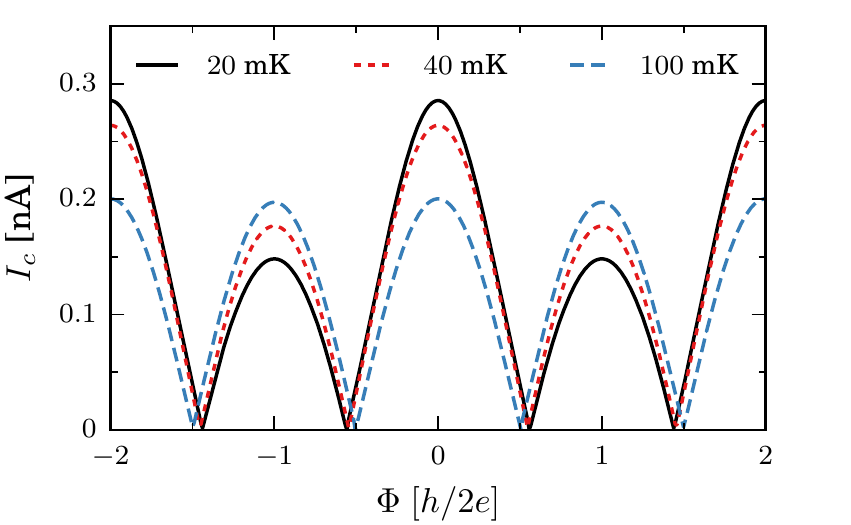}}
\caption{Fraunhofer oscillations for the experimentally relevant parameters given in the text, calculated from Eq.\ \eqref{IMatsubara} for three different temperatures.
}
\label{fig:experiment}
\end{figure}

\begin{figure}[tb]
\centerline{\includegraphics[width=0.9\linewidth]{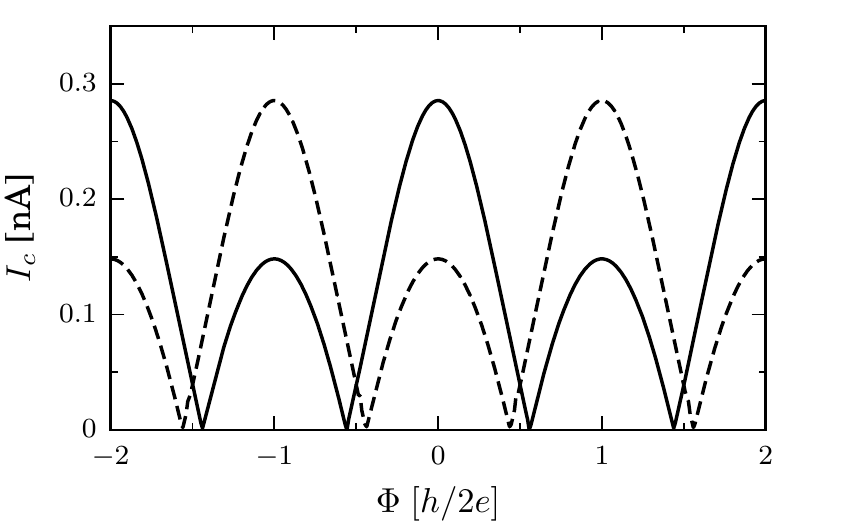}}
\caption{The solid curve is the $T=20\,{\rm mK}$ critical current of Fig.\ \ref{fig:experiment}, without phase shifts at the scattering nodes, while the dashed curve shows the inverted even-odd effect for $\psi_1'+\psi_3'=\pi$ (and all other phase shifts kept at zero).
}
\label{fig:phase}
\end{figure}

The InAs/GaSb quantum well with Ti/Al electrodes of Ref.\ \onlinecite{Pri14} has superconducting gap $\Delta_0 = 0.125\,{\rm meV}$ and edge state velocity\cite{Wan14} $v = 4.6 \cdot 10^4\,{\rm m/s}$. We take the same $v$ for the non-helical channel. There is some uncertainty in the effective dimensions of the junction, we set $L=0.5\,\mu{\rm m}$, $W=3.5\,\mu{\rm m}$. We then have comparable $L$ and $\xi_0=\hbar v/\Delta_0$, so we calculate the supercurrent directly from Eq.\ \eqref{IMatsubara} --- without taking the short-junction limit. The observed critical current in the $0.25\,{\rm nA}$ range implies an Andreev reflection probability $\Gamma\approx 0.2$, which is the value we take for $\Gamma_n$ at all four scattering nodes.

The degree of spin mixing upon propagation along the nonhelical channel is quantified by setting $U_1 U_3^\dagger=U_2 U_4^\dagger= e^{i\gamma\sigma_x}$. The value of $\gamma$ is unknown, we take a moderately strong spin mixing with $\gamma=\pi/6$, but note that the even-odd effect exists also without any spin mixing (see Fig.\ \ref{fig_evenodd}). The critical current shown in Fig.\ \ref{fig:experiment} exhibits an even-odd effect of a similar magnitude as observed experimentally.\cite{Pri14} The temperature dependence is somewhat stronger: In the experiment traces of the even-odd effect are still visible at $100~{\rm mK}$, but not in our calculation.

The beating mechanism has one qualitative feature that can help to distinguish it from other explanations of the even-odd effect: The sign of the effect --- whether the $\Phi=0$ peak is larger or smaller than the $\Phi=h/2e$ peak --- depends on microscopic details. This is evident from Eq.\ \eqref{Iresult1}, in that the offset $f$ can be of either sign. A similar inversion of the even-odd effect can be induced by varying the phase shifts in the node scattering matrix \eqref{snUn}, as we show in Fig.\ \ref{fig:phase}. Observation of an even-odd effect with the smallest peak at even multiples of $h/2e$ would constitute strong support for the beating mechanism, but no such inversion has been found so far.\cite{Pri14}

In our analysis we have assumed helical edge state transport, appropriate for a quantum spin-Hall insulator, but the beating mechanism itself would apply also to nonhelical edge conduction. As was also pointed out in the experimental paper,\cite{Pri14} the Fraunhofer oscillations are a sensitive probe of the current distribution, but cannot distinguish between a topologically trivial or nontrivial Josephson junction. That would require observation of a quantized conductance or supercurrent.

In conclusion, we have analyzed the effect of inter-edge coupling on the Fraunhofer oscillations in a quantum spin-Hall Josephson junction. A  network model allows for an efficient description of the beating of $h/2e$ periodic intra-edge and $h/e$ periodic inter-edge contributions to the critical current. The even-odd effect has comparable magnitude to what is observed in a recent experiment,\cite{Pri14} see Fig.\ \ref{fig:experiment}, but the sample-dependent inversion of Fig.\ \ref{fig:phase} has not been observed. 

We note that the beating mechanism studied here in the two-dimensional geometry of a quantum spin-Hall insulator may apply more generally when a pair of conducting pathways enclosing different flux interferes. Indeed, a recent work studies a similar beating effect in a one-dimensional wire geometry,\cite{Mir14} to explain multi-periodic Fraunhofer oscillations observed in Bi nanowires.\cite{Li14}

We acknowledge discussions on the experiment with L. P. Kouwenhoven and on the role of fermion-parity switches with I. C. Fulga, and we thank A. R. Akhmerov for a critical reading of the manuscript. This research was supported by the Foundation for Fundamental Research on Matter (FOM), the Netherlands Organization for Scientific Research (NWO/OCW), and an ERC Synergy Grant.

\appendix

\section{Network model of a Josephson junction}

We describe in more detail the network model of a Josephson junction that we have introduced and applied in the main text, and in particular give a selfcontained derivation of the formula \eqref{rhodef} for the supercurrent through the network.

\subsection{Construction of node and bond scattering matrices}

The scattering theory of a Josephson junction developed in Ref.\ \onlinecite{Bee91} expresses the supercurrent in terms of the two scattering matrices $s_{\rm N}$ of the normal region (N) and $s_{\rm A}$ of Andreev reflection at the normal-superconductor (NS) interfaces. While the matrix $s_{\rm A}$ has a simple expression, see Eq.\ \eqref{sAdef}, calculation of $s_{\rm N}$ can be quite complicated. 

\begin{figure}[tb]
\centerline{\includegraphics[width=0.8\linewidth]{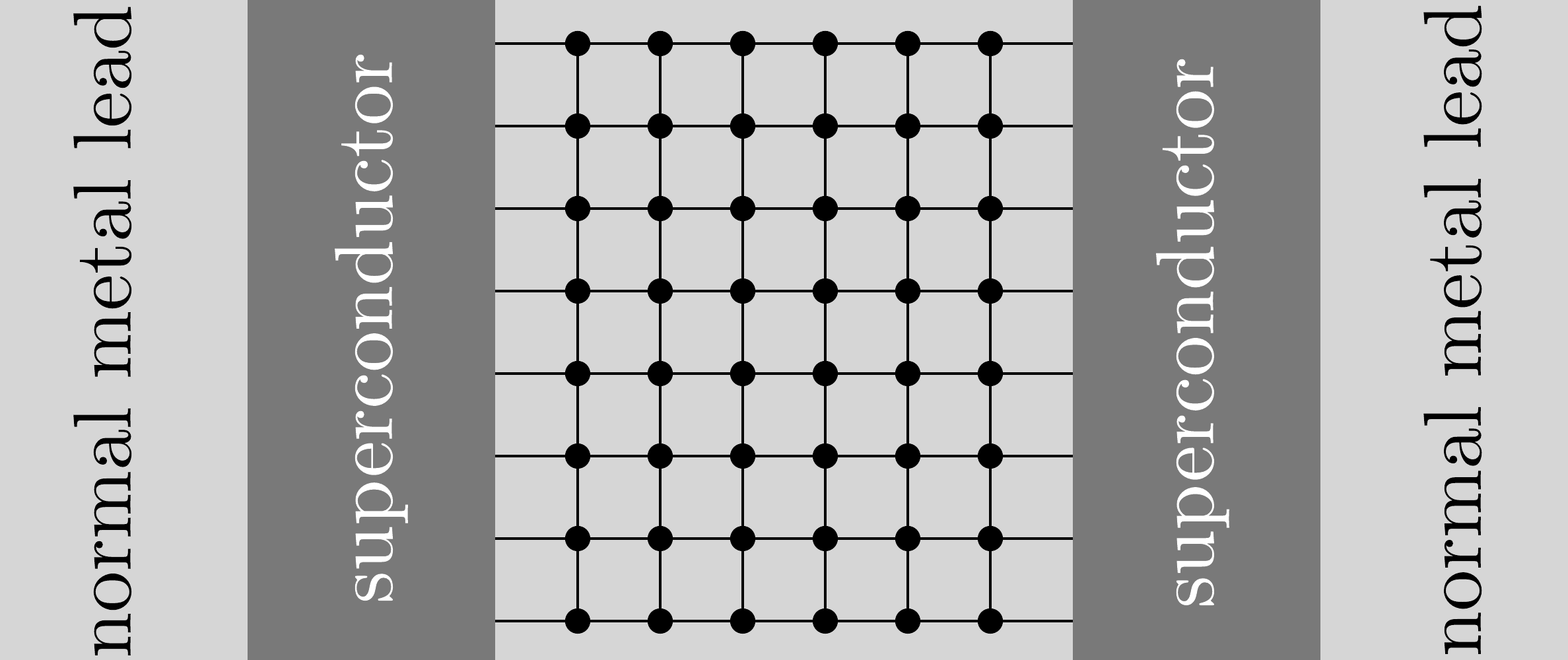}}
\caption{Network model of a Josephson junction. The normal metal leads are attached to the superconductor as an intermediate step in the derivation of the scattering matrix formula for the supercurrent. The final expression \eqref{Inoparity} contains only the scattering matrices of the nodes and bonds in the junction region. Andreev reflection at the interface with the superconductor is included in the bond matrix, via Eq.\ \eqref{alphadef2}.
}
\label{network}
\end{figure}

In this work we have used an alternative network representation, where the supercurrent is expressed in terms of the scattering matrices $s_{\rm node}$ and $s_{\rm bond}$ of the nodes and bonds of a network (see Fig.\ \ref{network}).  These matrices are the direct sum of scattering matrices of individual nodes and bonds, so they have a simple structure that can be written down without any calculation. 

The node matrix $s_{\rm node}$ is block-diagonal with the scattering matrices $s_{n}$ of node $n=1,2,\ldots$ on the diagonal. Because electrons and hole are uncoupled in the normal region, each matrix $s_{n}$ is itself block-diagonal with an electron block $s_{n,e}(\varepsilon)$ and a hole block $s_{n,h}(\varepsilon)=s_{n,e}^\ast(-\varepsilon)$. We thus have $s_{\rm node}=s_{1e}\oplus s_{1h}\oplus s_{2e}\oplus s_{2h}\oplus\cdots$.

The bond matrix $s_{\rm bond}=UP$ is the product of a diagonal matrix $U$ of phase factors and a permutation matrix $P$ that maps the indices of outgoing modes to incoming modes. The mode indices are spin $s\in\{\uparrow,\downarrow\}\equiv\{+1,-1\}$, particle-hole $t\in\{e,h\}\equiv\{+1,-1\}$, and possibly also an orbital degree of freedom $\nu\in\{1,2,\ldots\}$. (In the system considered in the main text all bonds support only a single orbital mode.) The matrix element $\langle n\nu' s't'|s_{\rm bond}|m\nu st\rangle$ is zero unless a mode with spin $s$ of particle-hole type $t$ that is outgoing from node $m$ in orbital mode $\nu$ is incoming onto node $n$ in orbital mode $\nu'$ as a spin-$s'$ type-$t'$ particle. There are no ``dangling bonds'', meaning that $s_{\rm bond}$ has a single non-zero element in each row and column.

Andreev reflection at the NS interface is included in $s_{\rm bond}$ via the matrix elements
\begin{equation}
\begin{split}
&\langle n\nu' s't'|s_{\rm bond}|m\nu st\rangle=-i\alpha\,st\,\delta_{mn}\delta_{\nu\nu'}\delta_{s',-s}\delta_{t',-t},\\
&\alpha(\varepsilon)=ie^{-i\,{\rm arccos}(\varepsilon/\Delta_0)}=i\varepsilon/\Delta_0+\sqrt{1-\varepsilon^2/\Delta_0^2}.
\end{split}
\label{alphadef2}
\end{equation}
Please note that this definition of $\alpha$ differs by a factor $i$ with that used in Ref.\ \onlinecite{Bee91}; we prefer it this way because now $\alpha(\varepsilon)=\alpha^\ast(-\varepsilon)$, so it is particle-hole symmetric. The branch of the square root of $1-\varepsilon^2/\Delta_0^2$ is fixed by ${\rm Re}\,\alpha(\varepsilon+i0^+)>0$, so that for $|\varepsilon|>\Delta_0$ one has
$\alpha = i\varepsilon/\Delta_0-i({\rm sign}\,\varepsilon)\sqrt{\varepsilon^2/\Delta_0^2-1}$. 

For $|\varepsilon|<\Delta_0$ one has $|\alpha|=1$, hence Eq.\ \eqref{alphadef2} describes Andreev reflection with unit probability. This is a matter of convenience, because a nonzero probability of normal reflection at the NS interface can be accounted for by inserting a node just before the interface. (See Ref.\ \onlinecite{Giu13} for an alternative scattering formulation that does not separate normal and Andreev reflection.)

The simplification afforded by the network representation in the construction of the scattering matrices comes at a price: The matrix $s_{\rm node}s_{\rm bond}$ is sparse, but its dimension is much larger than the dimension of $s_{\rm N}s_{\rm A}$. We have not studied this systematically, but we expect both representations in terms of $s_{\rm node}s_{\rm bond}$ and $s_{\rm N}s_{\rm A}$ to have the same computational complexity, scaling $\propto N^3$ with the number of nodes.

\subsection{Density of states in terms of node and bond matrices}

To calculate the density of states of the Josephson junction it is convenient to attach normal metal leads to the superconductors (see Fig.\ \ref{network}). The leads support the propagating modes that form basis states for the scattering matrix $S_{\rm SNS}(\varepsilon)$ of the junction. (Without the normal leads we would only have propagating modes above the gap, for $|\varepsilon|>\Delta_0$.) 

The density of states $\rho(\varepsilon)$ is determined by the unitary scattering matrix $S_{\rm SNS}$ via the general expression\cite{Akk91}
\begin{equation}
\rho(\varepsilon)=\frac{1}{2\pi}\frac{d}{d\varepsilon}\,{\rm Im}\,\ln\,{\rm Det}\,S_{\rm SNS}(\varepsilon+i0^+).\label{rhoDetS}
\end{equation}
In Ref.\ \onlinecite{Bee91} the determinant of $S_{\rm SNS}$ was related to the determinant of $1-s_{\rm N}s_{\rm A}$. Here we seek to derive a similar expression in terms of the node and bond matrices of the network.

For $|\varepsilon|<\Delta_0$ the bond matrix $s_{\rm bond}(\varepsilon)$ is unitary, but for $|\varepsilon|>\Delta_0$ the Andreev reflection probability $|\alpha|^2$ drops below unity because of propagating modes in the superconductor. Unitarity can be restored by embedding $s_{\rm bond}$ in larger matrix
\begin{equation}
S_{\rm bond}=\begin{pmatrix}
s_{\rm bond}&t_{\rm NS}\\
t'_{\rm NS}&r_{\rm NS}
\end{pmatrix},\label{Sbond}
\end{equation}
containing also the transmission and reflection matrices of the NS interfaces: a mode incident from the normal lead onto the NS interface is reflected with amplitude $r_{\rm NS}$ and is transmitted through the interface with amplitude $t_{\rm NS}$, while $t'_{\rm NS}$ describes the transmission in the opposite direction (into the normal lead). At subgap energies $t_{\rm NS}=t'_{\rm NS}=0$, while $r_{\rm NS}$ as well as $s_{\rm bond}$ are unitary. Above the gap only the full matrix $S_{\rm bond}$ is unitary.

In order to rewrite Eq.\ \eqref{rhoDetS} in terms of $s_{\rm node}$ and $s_{\rm bond}$ we start from the relation
\begin{align}
S_{\rm SNS}&=r_{\rm NS}+\sum_{n=0}^{\infty}t'_{\rm NS}s_{\rm node}(s_{\rm bond}s_{\rm node})^{n}t_{\rm NS}\nonumber\\
&=r_{\rm NS}+t'_{\rm NS}s_{\rm node}(1-s_{\rm bond}s_{\rm node})^{-1}t_{\rm NS}\nonumber\\
&=r_{\rm NS}-t'_{\rm NS}(s_{\rm bond}-s_{\rm node}^\dagger)^{-1}t_{\rm NS}.\label{Ssrelation}
\end{align}
This relation expresses the fact that modes incident on the SNS junction are either reflected directly at the NS interface, with amplitude $r_{\rm NS}$, or first transmitted through the interface (amplitude $t_{\rm NS}$), followed by multiple scattering in the network (amplitude $s_{\rm node}+s_{\rm node}s_{\rm bond}s_{\rm node}+\cdots$), and finally transmission back through the NS interface (amplitude $t'_{\rm NS}$). In the final equality in Eq.\ \eqref{Ssrelation} we have used that $s_{\rm node}$ (unlike $s_{\rm bond}$) is unitary for all energies.

We now invoke the folding identity,
\begin{equation}
{\rm Det}\,\begin{pmatrix}
A&B\\
C&D
\end{pmatrix}=({\rm Det}\,A)\,{\rm Det}\,(D-CA^{-1}B),
\end{equation}
valid for any invertible submatrix $A$, to equate
\begin{align}
&{\rm Det}\,(s_{\rm bond}-s_{\rm node}^\dagger)\,{\rm Det}\,S_{\rm SNS}\nonumber\\
&\quad={\rm Det}\,\begin{pmatrix}
s_{\rm bond}-s_{\rm node}^\dagger&t_{\rm NS}\\
t'_{\rm NS}&r_{\rm NS}
\end{pmatrix}\nonumber\\
&\quad={\rm Det}\,\left[S_{\rm bond}-\begin{pmatrix}
s_{\rm node}^\dagger&0\\
0&0
\end{pmatrix}\right]\nonumber\\
&\quad={\rm Det}\,S_{\rm bond}\,{\rm Det}\,\left[1-\begin{pmatrix}
s_{\rm node}^\dagger&0\\
0&0
\end{pmatrix}S_{\rm bond}^\dagger\right]\nonumber\\
&\quad={\rm Det}\,S_{\rm bond}\,{\rm Det}\,(1-s_{\rm node}^{\dagger}s_{\rm bond}^\dagger)\label{Ssrelation2}
\end{align}
\begin{align}
&\Rightarrow{\rm Det}\,S_{\rm SNS}={\rm Det}\,S_{\rm bond}\,{\rm Det}\,s_{\rm node}\frac{{\rm Det}\,(1-s_{\rm node}^\dagger s_{\rm bond}^\dagger)}{{\rm Det}\,(1-s_{\rm  node}s_{\rm bond})}\nonumber\\
&\quad=\frac{({\rm Det}\,S_{\rm bond}\,{\rm Det}\,s_{\rm node})^{1/2}\,{\rm Det}\,(1-s_{\rm node}^\dagger s_{\rm bond}^\dagger)}{({\rm Det}\,S_{\rm bond}^\dagger\,{\rm Det}\,s_{\rm node}^\dagger)^{1/2}\,{\rm Det}\,(1-s_{\rm node} s_{\rm bond})}.\label{Ssrelation3}
\end{align}
In the final equality we have used that both $S_{\rm bond}$ and $s_{\rm node}$ are unitary.

The folding identity also tells us that
\begin{align}
&{\rm Det}\,S_{\rm bond}={\rm Det}\,s_{\rm bond}\,{\rm Det}\,s_{\rm lead},\label{DetSbond}\\
&s_{\rm lead}=r_{\rm NS}-t'_{\rm NS}s_{\rm bond}^{-1}t_{\rm NS},\label{slead}
\end{align}
where $s_{\rm lead}$ describes the reflection of a mode incident from the normal metal lead when all bonds of the network are cut at the first node from the NS interface.  We can therefore identify
\begin{equation}
\rho_{\rm lead}(\varepsilon)=\frac{1}{2\pi }\frac{d}{d\varepsilon}\,{\rm Im}\,\ln\,{\rm Det}\,s_{\rm lead}(\varepsilon+i0^+)\label{rholead}
\end{equation}
with the density of states of the SNS junction without the normal region.

Combination of Eq.\ \eqref{rhoDetS} with Eqs.\ \eqref{Ssrelation3} and \eqref{slead} gives the required scattering formula for the density of states of the Josephson junction,
\begin{subequations}
\label{rhoresult}
\begin{align}
&\rho(\varepsilon)={\rm Im}\,\frac{d}{d\varepsilon}\nu(\varepsilon+i0^+)+\rho_{\rm lead}(\varepsilon),\label{rhoresulta}\\
&\nu(\varepsilon)=-\pi^{-1}\ln\,{\rm Det}\,(1-s_{\rm node} s_{\rm bond})\nonumber\\
&\qquad\quad+\tfrac{1}{2}\pi^{-1}\ln\,{\rm Det}\,(s_{\rm node} s_{\rm bond}).\label{rhoresultb}
\end{align}
\end{subequations} 
This is Eq.\ \eqref{rhodef} from the main text, where the $\phi$-independent terms $\rho_{\rm lead}$ and $\ln{\rm Det}\,s_{\rm node}s_{\rm bond}$ are simply referred to as ``constant''. The formula describes both the discrete and the continuous spectrum: For $|\varepsilon|<\Delta_0$ it gives a sum of delta functions at the bound state energies, superimposed on the smooth $\rho_{\rm lead}$, while for $|\varepsilon|>\Delta_0$ these peaks are broadened because the bound states can leak out into the superconductor. 

\subsection{Supercurrent in terms of node and bond matrices}

In the absence of fermion parity conservation (the case treated in the main text) we need to only retain the $\phi$-independent term $\propto\ln{\rm Det}\,(1-s_{\rm node}s_{\rm bond})$ in the density of states \eqref{rhoresult}. As derived in Ref.\ \onlinecite{Bro97}, the supercurrent at temperature $T$ is then a sum of the logarithmic determinant over fermionic Matsubara frequencies $\omega_p=(2p+1)\pi kT$,
\begin{align}
&I_{0}=-kT\frac{2e}{\hbar}\frac{d }{d \phi}\sum_{p=0}^\infty\ln {\rm Det}\, [1-s_{\rm node}(i\omega_p)s_{\rm bond}(i\omega_p)]\nonumber\\
&=kT\frac{2e}{\hbar}\sum_{p=0}^\infty\,{\rm Tr}\left\{[1-s_{\rm node}s_{\rm bond}]^{-1}s_{\rm node}\frac{ds_{\rm bond}}{d\phi}\right\}_{\varepsilon=i\omega_p}.
\label{Inoparity}
\end{align}
At zero temperature the sum may be approximated by an integral, $kT\sum_p\mapsto \int_0^\infty d\omega/2\pi$. The factor of $2e$ refers to the Cooper pair charge. Ref.\ \onlinecite{Bro97} has an additional factor of two because of spin degeneracy, which here we do not assume.

The derivation of Eq.\ \eqref{Inoparity} in Ref.\ \onlinecite{Bro97} was for ${\rm Det}\,(1-s_{\rm N} s_{\rm A})$, but it holds equally well for ${\rm Det}\,(1-s_{\rm node}s_{\rm bond})$ because it only relies on two properties of $\nu$ that are universally valid: particle-hole symmetry, $\nu(\varepsilon)=\nu^\ast(-\varepsilon)$, and causality --- $\nu(\varepsilon)$ being an analytic function for ${\rm Im}\,\varepsilon>0$. 

When fermion parity is conserved the terms $\rho_{\rm lead}$ and $\ln{\rm Det}\,(1-s_{\rm node}s_{\rm bond})$ in Eq.\ \eqref{rhoresult} must be retained even though they are not $\phi$-dependent, because they are needed to determine whether a set of occupation numbers has the right fermion parity. It is for this reason that we were careful to properly account for these $\phi$-independent terms in the calculation of the density of states. The expression for the supercurrent in the fermion-parity conserving case contains also a sum over bosonic Matsubara frequencies $\Omega_p=2p\pi kT$,
\begin{align}
I_\pm={}&I_0-kT\frac{2e}{\hbar}\frac{d }{d \phi}\ln\tfrac{1}{2}\left[1\pm e^{J_{\rm lead}}\sqrt{{\rm Det}\,X(0)}\right.\nonumber\\
&\times\left.\exp\left(\sum_{p=1}^{\infty}(-1)^{p}\ln{\rm Det}\,X(i\Omega_p/2)\right)\right],\label{Iparity}\\
X={}&(1-s_{\rm node}s_{\rm bond})(s_{\rm node}s_{\rm bond})^{-1/2},\label{Xdef}\\
J_{\rm lead}={}&\int_{0}^{\infty}d\varepsilon\,\rho_{\rm lead}(\varepsilon)\ln\tanh(\varepsilon/2kT).\label{Jleaddef}
\end{align}
The $\pm$ sign in Eq.\ \eqref{Iparity} indicates even or odd fermion parity of the superconducting ground state. The sign is + at $\phi=0$, and then switches each time a pair of bound states crosses the Fermi level ($\varepsilon=0$).

One limitation of the network representation is that we do not have a formula for the ground-state fermion parity in terms of $s_{\rm node}$ and $s_{\rm bond}$. The derivation in Ref.\ \onlinecite{Bee13} of such a formula in terms of $s_{\rm N}$ and $s_{\rm A}$ does not carry over. 


\begin{thebibliography}{99}
\bibitem{appendix} The derivation of Eq.\ \eqref{rhodef} is given in the Appendix.
\end{thebibliography}
\end{document}